\documentclass[conference]{IEEEtran}
\usepackage{amsmath,amssymb,verbatim}
\usepackage{graphicx}
\usepackage{amssymb}
\usepackage{epstopdf}

\newtheorem{definition}{Definition}[section]
\newtheorem{thm}{Theorem}[section]
\newtheorem{proposition}[thm]{Proposition}

\newtheorem{exam}{Example}[section]

\newtheorem{remark}{Remark}[section]

\newcommand{\Z}{{\mathbf Z}}
\newcommand{\R}{{\mathbf R}}
\newcommand{\C}{{\mathbf C}}

\newcommand{\OO}{{\mathcal O}}

\newcommand{\A}{{\mathcal A}}

\newcommand{\Q}{{\mathbb Q}}

\begin{document}

\title{Space--Time Storage Codes for Wireless Distributed Storage Systems}

\author{
\IEEEauthorblockN{Camilla Hollanti, 
David Karpuk, and 
Amaro Barreal} 
\IEEEauthorblockA{Department of Mathematics and Systems Analysis\\ Aalto University, Finland\\ 
        Emails: firstname.lastname@aalto.fi}
        \and
        \IEEEauthorblockN{Hsiao-feng (Francis) Lu}
\IEEEauthorblockA{Department of Electrical and Computer Engineering\\ National Chiao Tung University,
        Hsinchu, Taiwan\\
        Email: francis@mail.nctu.edu.tw}
        }

\maketitle

%
%

\begin{abstract}
Distributed storage systems (DSSs) have gained a lot of interest recently, thanks to their robustness and scalability compared to single-device storage.  Majority of the related research has exclusively concerned the network layer. At the same time, the number of users of, \emph{e.g.}, peer-to-peer (p2p) and device-to-device (d2d) networks as well as  proximity based services is growing rapidly, and the mobility of users is considered more and more important. This motivates, in contrast to the existing literature, the study of the physical layer functionality of wireless distributed storage systems. 

In this paper, we take the first step towards protecting the storage repair transmissions from physical layer errors when the transmission takes place over a fading channel. To this end, we introduce the notion of a \emph{space--time storage code}, drawing together the aspects of network layer and physical layer functionality and resulting in cross-layer robustness. It is also pointed out that existing space--time codes are too complex to be utilized in storage networks when the number of helpers involved is larger than the number of receive antennas at the newcomer or data collector, hence creating a call for less complex transmission protocols. 

\end{abstract}


%
%

\section{Introduction}

Our society relies on wireless communications and data storage over unreliable channels and networks more than ever, largely due to increasing demand for wireless services,  social networks and different types of peer-to-peer (p2p) systems such as digital video broadcasting, Facebook, Google, Oracle, and various video-on-demand (VoD) services.   


Distributed storage systems (DSSs) enable reliable data storage by storing data on separate devices in a redundant way. The simplest form of distributed storage is that of replication: by storing some number of replicas of the original file on the devices in the storage network, one is able to maintain and retrieve the data even when some of the nodes fail, provided at least one of them survives. Node failure is common, and can be caused by hardware corruption, system overload (\emph{e.g.} in p2p systems), or by a node leaving the system.  

If the data is stored over $n$ storage nodes by using an $(n,k)$ maximum distance separable (MDS) code, the whole data file can be reconstructed by contacting any $k$ out of $n$ nodes. In addition to storing the file, the system has to be repaired by replacing a node with a new one whenever some node fails. This can be done by using, \emph{e.g.}, \emph{regenerating codes} \cite{DGWWR07}. If the \emph{newcomer} node replacing the failed node has to contact  $d$ \emph{helper} nodes in order to restore the contents of the lost node, we call the code an $(n,k,d)$ code. Recent work \cite{DGWWR07,me} considers tradeoffs between the storage capacity, secrecy capacity, and repair bandwidth. Explicit storage code constructions achieving some of the tradeoffs can be found in \cite{RR10, DRWS10, rashmietal}, among many others. Regenerating codes by definition achieve the storage capacity--repair bandwidth tradeoff.

In this paper, we will consider the physical layer functionality of distributed storage systems, where the storage nodes are scattered in a wireless network, or data lying in a (wired) data center is retrieved over a wireless access network. When a data collector requests a file, or a newcomer requests help from some survivor nodes in order to replace a lost node, the data transmissions that follow will thus take place over a wireless fading channel\footnote{
 The authors are aware that there is a long way from the application layer to the physical layer, and the original stored file will be chopped into packets with headers more than once when traveling through the different intermediate layers. Hence, instead of fractions of the `file' would be more appropriate to consider (fractions of) packets to be stored on and transmitted by the nodes in a wireless network. Nevertheless, to maintain the intuition, we have chosen to use the term `file', albeit admittedly somewhat inaccurate.}. Being able to perform transmission requests, in principle, requires a feedback (uplink) channel or a base station performing these requests. In the present paper, we will ignore this aspect and only concentrate on the protection of the downlink transmissions, \emph{i.e.}, of the transmissions from the storage nodes to a data collector or to a newcomer.

\subsection{Contributions and related work}
Most storage-related research focuses on the (logical) network layer, while the physical layer functionality is usually ignored due to the fact that many storage systems in big data centers are wired. Nonetheless, the authors are aware of some interesting works considering the physical layer. In \cite{gong}, a so-called partial downloading scheme is proposed that allows for data reconstruction with limited bandwidth by downloading only parts of the contents of the helper nodes. In \cite{Ning}, the use of a forward error correction code (\emph{e.g.}, LDPC code) is proposed in order to correct bit errors caused by fading. In \cite{rashmi_erasure}, optimal storage codes are constructed for the error and erasure scenario.  
The present paper deviates from the previous work in that it addresses the actual \emph{encoding} of the transmitted repair data in order to fight the effects caused by fading.


Isolated from the storage point of view, on the other hand,  a plethora of research has been carried out during the past two decades  in wireless communications (see \cite{oggierviterbo2} and the references therein). Motivated by this work, we will introduce the notion of  \emph{space--time storage codes}, a class of codes that should be able to resist fading of the signals during repair transmissions, while also maintaining the repair property of the underlying storage code.

The contributions of this paper are listed below.

\begin{itemize}
\item For the first time, protecting a distributed storage system from physical layer errors is considered in conjunction with the encoding of the repair and reconstruction transmissions in order to overcome the defect caused by fading. 
\item It is pointed out that the data reconstruction and repair transmissions can be modeled as a multiple access channel (MAC), providing us with a rich theory of MAC systems to be harnessed in the context of data storage. Further, a joint design of a storage code and a MAC space--time code, referred to as a \emph{space--time storage code}, is proposed as a protocol for wireless storage transmissions. 
\item Tentative design criteria for such codes are proposed based on earlier work on MAC space--time codes. 
\item Simulations are carried out to confirm the merit of the proposed protocol. 
\item Open questions and some problems are addressed to motivate further research. 
\end{itemize}

\section{Space--time storage codes}
Space--time coding has gathered an enormous amount of interest during the past 15 years due to many practical applications, among which are mobile phones, digital video broadcasting, satellite communications, and multiple access channels.  Especially algebraic number theory has offered a wide range of tools for packing the information into vectors and further into matrices in a robust and efficient way to reduce the required transmission power and the error probability \cite{oggierviterbo2}.

 A \emph{space--time (ST) code}  is a finite subset of the space of  $n_t\times T$ complex matrices $M_{n_t\times T}(\C)$, where $n_t$ denotes the number of transmit antennas, and $T$ is the number of channel uses, also called decoding delay. In order to avoid accumulation among codewords, it is preferable to consider codes with a lattice structure. A \emph{lattice} $\Lambda$ is a discrete finitely generated abelian subgroup of some ambient space $V$, \emph{e.g.}, $V=M_{n_t\times T}(\C)$.
 
 \begin{definition}
 A \emph{space--time lattice code} $\mathcal{C}_{ST}$ is of the form
 $$
 \mathcal{C}_{ST}=\left\{\sum_{i=1}^{\mathrm{rank}(\Lambda)}z_iB_i\in \Lambda\ |\ z_i\in \mathcal{S}\subset\Z\right\},
 $$
 where $B_1,\ldots,B_{\mathrm{rank}(\Lambda)}$ is a lattice basis for some lattice $\Lambda\subset M_{n_t\times T}(\C)$, and $\mathcal{S}$ is a finite signaling alphabet, \emph{e.g.}, pulse amplitude modulation (PAM).
 \end{definition}
 
 We refer the reader to \cite{oggierviterbo2} for a general introduction to space--time lattice codes. 

\subsection{Equivalence of wireless storage transmissions and MAC}
In a noncooperative multiple access channel, multiple users are simultaneously communicating to a joint destination, hence the destination receives a combination of all the transmissions. Virtually, this can be considered as a MIMO space--time code satisfying certain properties. The key difference to single-user ST codes is that the transmissions of different users (corresponding to different (groups of) antennas in the single-user case) should be independent of each other, whereas in the single-user case the message is encoded over many (groups of) antennas to provide better diversity. In the case of data storage applications, the different storage nodes typically neither share the same storage contents nor cooperate,  and hence the protocol resembles the MAC case rather than a single-access channel.  Hence,  the following observation is immediate.

\begin{proposition}\label{mac-equality}
In a distributed storage system, the repair (resp. reconstruction) transmission over a wireless fading channel can be modeled as a multiple access channel. The number of MAC users $K$ corresponds to the number of helpers involved in the repair (resp. reconstruction) process $K=d$ (resp. $K=k$). Furthermore, the MAC transmission can be virtually modeled as  a multiple-input multiple-output (MIMO) transmission described by the channel equation 
$$Y=HX+W,$$
where $X$ is the overall transmitted matrix of the $K$ users, $H$ and $W$ are the random fading and noise matrices, and $Y$ is the received matrix. 
\end{proposition} 
We refer the reader to \cite{kuser} for more details.

Suppose now that a node fails, and that an incoming node $u_{new}$ has to contact $K\in\{d,k\}$ nodes for repair/reconstruction.  Let us write
$
u_{i_1},\ldots,u_{i_K}
$
for the nodes contacted by $u_{new}$.  Each $u_{i_j}$ would like to send its contents $\bar{x}_{i_j}$ to $u_{new}$ over a Rayleigh fading channel.  To do this, we incorporate into our coding strategy  a bijective lift function
\[
L:\mathcal{X}\rightarrow \mathcal{C}, 
\]
where $\mathcal{X}$ is the set of possible encoded\footnote{Encoded by an MDS or other erasure code.} file fragments, and $\mathcal{C}$ is a finite symbol set of  size equal to the size of $\mathcal{X}$. We define $L(\bar{x}_i) = x_i$. To be more precise, we define below what we mean by a space--time storage code.

\begin{definition}\label{ST-storage}
A \emph{space--time storage code} consists of the following:
\begin{enumerate}
\item a DSS system with parameters defined as above, employing an $(n,k)$ MDS code or some other type of storage code,
\item a symbol set $\mathcal{C}$ carved from  $\C^s$ or $\R^{2s}$,
\item a bijective lift function
$
L:\mathcal{X}\rightarrow \mathcal{C}, 
$
where $\mathcal{X}$ is the set of possible encoded file fragments, and
\item a MAC space--time transmission protocol using $\mathcal{C}$ as information symbols.
\end{enumerate}
\end{definition}

 We point out the generality of this definition: it does not address the details of the storage code nor the space-time code. Here, we will concentrate on algebraic space-time lattice codes.


 Now let us consider a distributed storage system employing
a storage code, \emph{e.g.}, a regenerating code, and where the storage nodes are scattered
in a wireless network. 
 The data collector and newcomers connect to the
helper nodes over a fading channel, and download data symbols
from these nodes. 

We assume a Rayleigh fading channel, with the channel equation
$$
y=\sum_{i=1}^K h_ix_i+w,
$$
where $x_i$ is the codeword transmitted by the $i$th helper node, $h_i$ is the corresponding channel gain, and $w$ is the noise at the receiver. Here $h_i\in\C^{n_r\times 1}$ are i.i.d. Rayleigh distributed random variables with zero mean, $x_i\in\C^{1\times T}$, and $i=1,\ldots, K$, where $K$ is the number of helper nodes, so either $K=d$ for repair, or $K=k$ for file reconstruction\footnote{For simplicity,  we will only talk about repair from now on. The reconstruction process is similar, except for the number of nodes contacted being $k$ instead of $d$.}. The above channel equation can be transformed into an equivalent MIMO channel (cf. Prop. \ref{mac-equality}) as
$$
Y=HX+W.
$$

\begin{figure}
$$\framebox{\includegraphics[width=5cm]{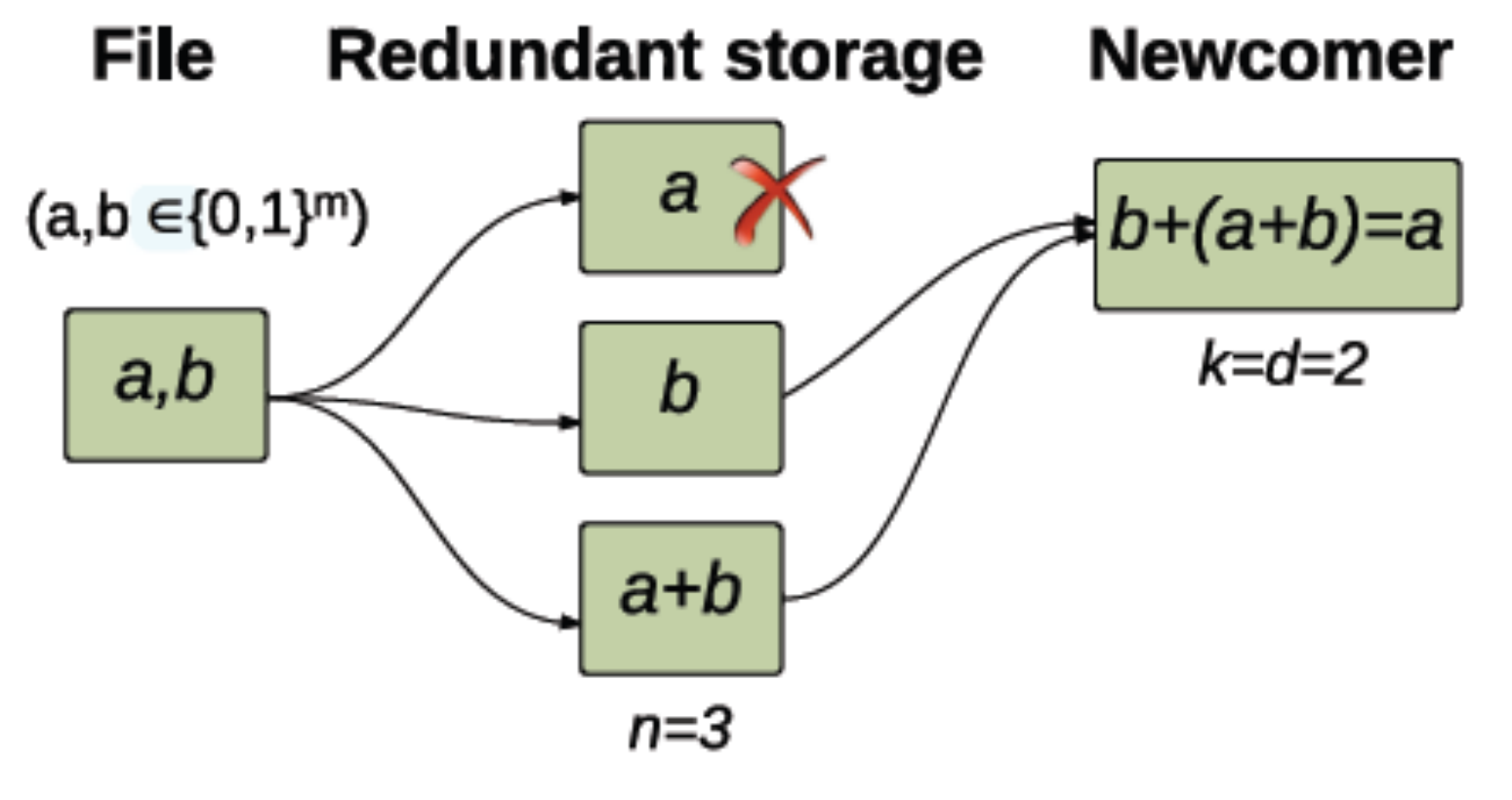}}$$
\caption{Repair process in a simple distributed storage system with $n=3$ and $k=d=2$ after the first node has failed.}
\label{toy}
 \end{figure}


\subsection{The lift function}
In a storage system, $a,b\in GF(2)^m=\{0,1\}^m$, or more generally $a,b\in GF(q)^m$, usually with $q=2^\ell$ for $\ell\geq 1$. The entries of a space--time code matrix, on the other hand,  are typically drawn from a complex alphabet $\subseteq \C$, \emph{e.g.}, from a ring of algebraic integers of a suitable number field. How should one deal with this potential conflict of different alphabets?
This is exactly where the lift function  $$L:GF(q)^m\rightarrow \mathcal{C}\subseteq \mathbf{C}$$ comes into the picture. 
There are various ways to design such a lift function, see \emph{e.g.} \cite{Lusina,Lu-Kumar}. The simplest option ($q=2$) is to map bit strings directly to PAM or quadrature amplitude modulation (QAM) alphabet via \emph{Gray}-mapping and then map the PAM or QAM symbols to algebraic integers via an integral basis over $\Z$ or $\Z[i]$, respectively.  Below, we describe yet another possibility to match bit strings to algebraic integers.

 Let $E/\Q$ be an imaginary quadratic field.  The ring of integers $\mathcal{O}_E$ of $E$ embeds as a lattice into $\C$ via the canonical embedding $\psi$.  
 Now let $\mathcal{I}_t=(2^t)$, an ideal in $\mathcal{O}_E$, which via the canonical embedding is a sublattice of $\mathcal{O}_E$.  Notice that we have a chain of inclusions:
\[
\cdots\subseteq \mathcal{I}_{t+1}\subseteq \mathcal{I}_{t} \subseteq \mathcal{I}_{t-1}\subseteq \cdots \subseteq \mathcal{I}_1 = (2) \subseteq \mathcal{O}_E 
\]
 There is a simple way  to put cosets of $\mathcal{I}_t$ in $\mathcal{O}_E$ in bijection with bit strings of length $2^{2t}$. The structure of the quotient ring $\mathcal{O}_E/\mathcal{I}_t$ depends on the factorization of the ideal $(2)$ in $\mathcal{O}_E$, which is easy to calculate given $E$.  We can write
\[
\{\text{bit strings of length $2^{2t}$}\}\longleftrightarrow \mathcal{O}_E/\mathcal{I}_t \longrightarrow \mathcal{O}_E \hookrightarrow \C\,,
\]
where the left-most arrow is the above outlined process, the middle arrow is choosing a coset representative, and the right-most arrow is the canonical embedding. Hence, we can identify the set of bit strings of length $2^{2t}$ with a finite subset of $\C$ of the same size.

 \section{Space--time storage codes over MAC}
 For simplicity, we start by a simple example (see Fig. \ref{toy}), which will be then  generalized in the end of this section.

 Assume we have mapped $a, b, c=a+b\in \{0,1\}^m$ to ring of integers in a number field $E$ as described above, and denote the resulting elements by $a',b',c'\in\mathcal{O}_E$.
 Assume the first node fails. The first helper, node 2, transmits the vector $\{b',\tau(b')\}$, while the second helper transmits  $\{c,\tau(c')\}$. Here $\tau$ is an   automorpism of $E$. 
 No collaboration between nodes required, and it is not even necessary for them to know who the other helpers are.

 The receiver observes 
 $$Y=H
\begin{pmatrix}
b'& \tau(b')\\
c' & \tau(c')\\
\end{pmatrix}+W.$$
 This corresponds to a MIMO MAC space--time code, which are known (\cite{kuser,remarks}) to perform well and to achieve the so-called diversity-multiplexing gain tradeoff (DMT) \cite{ZT-MAC}, when the field is chosen well. 
 With sufficient SNR, the receiver is able to decode the message with very high probability and gets $b',c'$, which he can map back to bit strings $b,c=a+b$ and further reveal $a=b+(a+b)$.

\begin{exam}\label{bb-example}
To provide an explicit example, we use the  following 2-user MAC ST-code  proposed in \cite{BB}. The code is based on the field extension $E=\Q(i,\sqrt{5})$ over $\Q(i)$. Let $\theta=\frac{1+\sqrt5}2$ and $\overline{\theta}=\frac{1-\sqrt5}2$. We denote by $\OO_E=\Z[i,\theta]$ the ring of integers of $E$, and by $(\alpha)=(1+i-i\overline{\theta})$ an ideal of $\OO_E$ used for constellation shaping.  Let $\tau$ denote the generator of the cyclic Galois group of $E/\Q(i)$ determined by $\theta\mapsto \overline{\theta}$. After mapping the bit strings in the three storage nodes into the elements $x_1,x_2,x_3\in\OO_E$ (\emph{e.g.}, as described above),  the code matrix used for the storage transmissions is 
$$
X=\begin{pmatrix}
\alpha x_{i_1}& \tau(\alpha) \tau(x_{i_1})\\
i\alpha x_{i_2} & \tau(\alpha)\tau(x_{i_2})\\
\end{pmatrix},
$$ 
where $i_1,i_2\in\{1,2,3\}$ and  the $i$th row of the matrix corresponds to the transmission by the $i$th helper node. In \cite{BB}, the authors further use  $i=\sqrt{-1}$ as a so-called twisting element in the lower left corner entry in order to get a full-rank matrix, which we have also added for consistency.  While it has been shown in \cite{kuser}  that this is not necessary for achieving the MAC DMT, the use of a twisting element may indeed be beneficial at low SNRs.
\end{exam}

Let us now consider the above setting in more generality. We denote by $b_1, \ldots, b_n$ the bit vectors stored at each node, with some redundancy (for instance, in our toy example we would have $b_3=b_1+b_2$). Let each node be occupied with $n_t$ transmit antennas. Assume in addition that the DSS uses an $(n,k,d)$ storage code.  Let $E\supseteq E_1 \supseteq E_2$ be a chain of cyclic Galois extension of degrees $[E:E_1]=2,\,[E_1:E_2]=d$ and $\tau$ the generator of the cyclic Galois group of $E_1/E_2$, and $\sigma$ the generator of the cyclic Galois group of $E/E_1$. Let $\mathcal{A}$ be a cyclic division algebra (CDA) with center $E_1$ and with a maximal subfield $E$, \emph{i.e.}, 
$$
\mathcal{A}=(E/E_1,\sigma,\gamma)=E\oplus eE\oplus\cdots\oplus e^{n_t-1}E,
$$
where $e^{n_t}=\gamma$,  $\gamma^i \notin N_{E/E_1}(E\setminus\{0\})$ for $i=1,\ldots, n_t-1$, and $x e=e\sigma(x)$ for all $x\in E$. See \cite{oggierviterbo2} for more details. 

Each element $a=x_0+\cdots+e^{n_t} x_{n_t},\, x_i\in E$, of $\A$ can be represented as a $n_t\times n_t$ matrix via the left regular representation (LRR) $\psi$.
Denote by  $x_{1j},\ldots, x_{n_tj},\, j=1,\ldots,d$ the elements corresponding to each of the $d$ helpers, obtained by mapping the bit strings $b_i$ to $\OO_E$, and by $X_{1},\ldots, X_{d}$ the matrices containing these elements obtained via the LRR. The MAC ST code matrix suitable for the repair transmission is described as follows:

\begin{equation}
\label{RST}
X=\begin{pmatrix}
 X_{1}&  \tau(X_{1})&\cdots &\tau^{d-1}(X_{1})\\
\vdots&&&\vdots\\
 X_{d} & \tau(X_{d}) & \cdots &\tau^{d-1}(X_{d})\\
\end{pmatrix}.
\end{equation}
We refer to \cite{kuser} for more details on the construction of MAC-DMT optimal space--time codes. 

Now,  after successfully decoding $X$ from $Y=HX+W$, we can map the $x_{ij}$ back to the corresponding bit strings $b_j$. Then, the actual repair can be performed by using the repair rule of the assigned  storage code. We point out that decoding  by a linear decoder such as sphere decoder requires the receiver to have $d$ antennas. This is an evident drawback for the proposed scheme when $d$ is large.  

\section{Design criteria for space--time storage codes} 

`Successfully' above requires that the SNR experienced at the receiver is high enough. As the channel quality is imposed by nature, we will concentrate on  designing the space--time code as well as possible. To this end, the ST code $\mathcal{C}_{ST}$ consisting of a finite number of the matrices \eqref{RST} should have the following property that we recite from \cite{remarks}.


\begin{definition} If the minimum of the nonzero determinants of the matrices $X$ defined as above is bounded from below by a positive constant, \emph{i.e.}, 
\vspace{2pt}
\begin{center}$\textstyle      \min \atop \scriptstyle  X\in\mathcal{C}_{ST},\det(X)\neq 0$ $\textstyle \{|\det(X)|\}\geq\kappa>0,\atop $
\end{center}
\vspace{2pt}
we say that the code $\mathcal{C}_{ST}$ has the \emph{conditional non-vanishing determinant (CNVD) property}. 
\end{definition}
If $X$ is not a square matrix, we extend our definition to consider $\det(XX^\dag)$, where $X^\dag$ is the complex conjugate transpose of $X$.

In order to achieve good performance, all the submatrices $S^{(j)}(S^{(j)})^\dag$ consisting of $j\leq d$ helpers should  have CNVD (cf. \cite{kuser}), where 
$$S^{(j)}=
\begin{pmatrix}
X_{i_1}&\cdots & \tau^{d-1}(X_{i_1})\\
\vdots& & \vdots \\
X_{i_j}&\cdots & \tau^{d-1}(X_{i_j})\\
\end{pmatrix}
$$
corresponds to any subset of $j$ helpers, $j=1,...,d$.

Let now $d$ denote the number of helpers all equipped with $n_t$ transmit antennas, and let $\A$ be a $E_1$-central algebra. The following theorem  \cite{CNVD} is the key to achieving the CNVD, but may impose restrictions on the parameters $d$ and $n_t$ and on the mapping $\tau$.

\begin{thm}[Center argument \cite{CNVD}] Let  $\A=\mathcal{M}_d(\frak{D})$ be a finite dimensional simple algebra, where $\frak{D}$ is an index $n_t$ division algebra. Then $\A$ is central simple over its center $E_1$, and the center is the same for $\A$ as for $\frak{D}$. The norm of an element $X$ of the matrix algebra $\A$ is the determinant of the matrix $X$.  
\end{thm}

 Hence, $\det(X) \in E_1$, and further $\det(X) \in \OO_{E_1}$, when we are using an $\OO_{E_1}$-order $\Lambda\subseteq\frak{D}$. If $E_1$ is either $\Q$ or $\Q(\sqrt{-m})$, we get  $|\det(X)|\geq 1$ \textbf{whenever it is nonzero}.

In addition to the ST code having a CNVD, the underlying storage code should be designed in such a way that 
 the probability of successful repair given the result $\hat{X}$ of the decoding of the received signal $Y=HX+W$, is maximized.  In the next section we will analyze the probability of successful repair in some example cases. 
 
\begin{remark} The code in Ex. \ref{bb-example} has CNVD if the twisting element $i$ is removed. Removing it does not affect the simulation results. The code is  MAC-DMT optimal without the twisting element \cite{kuser}, and we believe it is optimal  also with the twisting element. 
\end{remark}

 \section{Simulation results}
 
We have proposed to use MAC ST storage codes for wireless repair transmissions. In order to justify our proposition, we compare the repair bit errors of MAT ST storage codes to uncoded repair transmission carried out by simple (virtual) single or double spatial multiplexing, as explained in detail below. 

The plots in Figure 2 represent the slow fading scenario, assuming that the repair fragment size is 4 bits.  Simulations were carried out to investigate the repair of a failed node for the storage code in our toy example case (cf. Fig. \ref{toy}), where communication between the nodes takes place in a fading environment.  Each helper node transmits their assigned  bit string $b_i$ of 4 bits, and the bitwise XOR $\hat{b} = \hat{b}_1 + \hat{b}_2$ is calculated after decoding the received signal, where we denote the decoding outcome by $\hat{b}_1$ and $\hat{b}_2$. That is, if the decoding was successful, $b_1=\hat{b}_1, b_2=\hat{b}_2$.  Each of $b_1$ and $b_2$ is modulated according to the particular coding strategy used. We assume two antennas at the receiver for each scheme. 



\textbf{\emph{Double Spatial Multiplexing (DSM):}} Each of the $b_i$, $i=1,2$ is split in half as $b_i = b_{i1}b_{i2}$, and each substring of two bits is modulated into a 4-QAM symbol using the \emph{Gray}-mapping $g$, working as our lift function (cf. Def. \ref{ST-storage}). We define $x_{ij} := g(b_{ij})$ for $i=1,2$ and $j=1,2$.  Node $i$ then transmits $x_{i1}$ over the first channel use, and $x_{i2}$ over the second channel use.  The virtual MIMO channel is described by the equation
$
Y = HX + W,
$
where  $$X = \left(\begin{array}{cc} x_{11} & x_{12} \\ x_{21} & x_{22} \end{array}\right),
$$
$H$ is the $2\times 2$ channel matrix, $W$ is the $2\times 2$ noise matrix, and $Y$ is the received matrix.  Maximum likelihood decoding is performed to calculate
$
\hat{X} = \text{arg}\min_{X'}||Y-HX'||^2
$
where $X'$ ranges over all possible $X$ of the above form assuming 4-QAM, and $||\cdot||$ is the Frobenius norm.  We thereby obtain estimates $\hat{x}_{ij}$, to which $g^{-1}$ can be applied to obtain
$
\hat{b} = \hat{b}_1 + \hat{b}_2 = \hat{b}_{11}\hat{b}_{12} + \hat{b}_{21}\hat{b}_{22}
$
where $\hat{b}_{ij} = g^{-1}(\hat{x}_{ij})$, the reconstituted fragment.  A bit error occurs whenever one of the bits of $b$ and $\hat{b}$ differ. 

 \textbf{\emph{MAC Storage Code:}}  The setup is exactly the same as in the DSM case, except the 4-QAM symbols are further encoded using the MAC code described in Ex. \ref{bb-example}.

\textbf{\emph{Single Spatial Multiplexing (SSM):}} Each of the $b_i$, $i=1,2$ is modulated into a $16$-QAM symbol using the \emph{Gray}-mapping $g$.  We define $x_i = g(b_i)$ for $i=1,2$.  Node $i$ then transmits $x_i$ over the channel.  Only one channel use is needed to reconstruct the lost fragment.
 Maximum likelihood decoding is again performed to obtain $\hat{X}$ and thereby $\hat{x}_i$ for $i=1,2$.  The reconstructed file fragment is
$
\hat{b} = \hat{b}_1 + \hat{b}_2,\quad \hat{b}_i = g^{-1}(\hat{x}_i)
$
and a bit error occurs whenever one of the bits of $b$ and $\hat{b}$ differ.

We should note that the three coding strategies do not have the same data rate.  In particular, The DSM and MAC Storage Code strategies transmit 4 bits per channel use (bpcu), while the SSM transmits 8 bpcu.  Thus the comparison is not between codes of the same rate, but rather transmission schemes for recovering file fragments of the same size.  One can see from the simulation results that the DSM and MAC Storage Code strategies have a lower BER, while taking two channel uses to repair.  However, the SSM strategy can recover the file in just one channel use, at the expense of requiring more energy for the same BER.

\begin{figure}\label{slow}
$$\includegraphics[width=6cm]{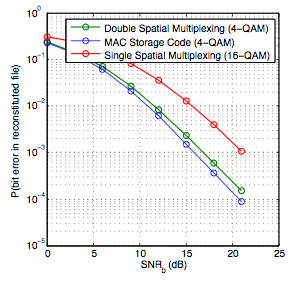}$$
\vspace{-1cm}
\caption{BER of ST storage codes for $2\times 2 $ slow fading MAC channel and fragment size of 4 bits (\emph{i.e.}, $a,\, b$, and $a+b$ are  bit strings of length 4).}
\end{figure}
\begin{figure}\label{fast}
$$ \includegraphics[width=6cm]{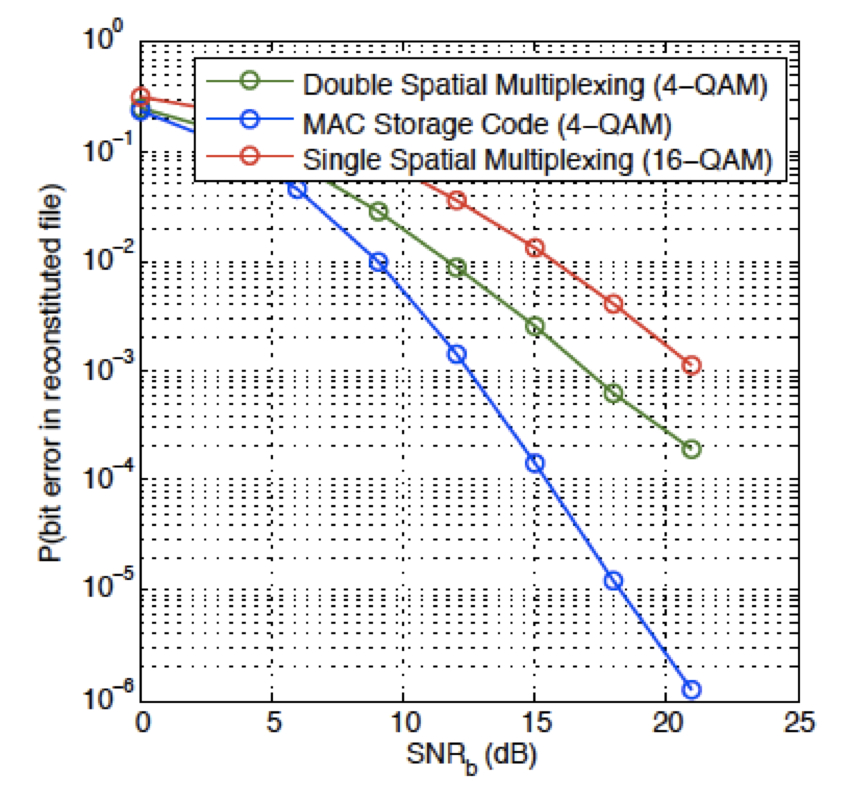}$$
\vspace{-1cm}
\caption{BER of ST storage codes for $2\times 2$  fast fading MAC channel and fragment size of 4 bits.}
\end{figure}


The plots in Figure 3 represent the fast fading scenario, where the channel change every channel use, and is independent of the previous channel state.  The fragment sizes and coding strategies are otherwise identical to the slow fading scenario. 

Since the SSM strategy requires only one channel use, its performance is the same over slow and fast fading channels.

\section{Discussion}

We have defined  space--time storage codes that are able to maintain and repair data that lies in storage systems operating over  wireless fading channels. Here, the focus was on embedding a storage code into a MAC space--time code, but what is potentially more interesting is the question as to how to jointly design a ST storage code from scratch such that probability that the system maintains its functionality is maximized. Studying this probability may give rise to new, more delicate design criteria for  ST storage codes, instead of just optimizing the storage code and the space--time code separately.  Combining optimal MAC ST codes and storage codes is problematic also due to high complexity: the MAC ST storage codes proposed in this paper require $K$ antennas at the receiver in order to perform sphere decoding when there are $K$ helper nodes. Hence, new repair transmission protocols with lower complexity are called for, while ideally maintaining good performance and achieving the DMT.


\begin{thebibliography}{1}


\bibitem{DGWWR07}
A.~Dimakis, P.~Godfrey, Y.~Wu, M.~Wainright, and K.~Ramchandran,
\newblock ``Network coding for distributed storage systems'',
\newblock {\em IEEE Trans. Inf. Theory}, vol. 56, no. 9, 
Sep. 2010.


\bibitem{RR10}
S.~{E}l Rouayheb and K.~Ramchandran,
\newblock ``Fractional repetition codes for repair in distributed storage
  systems'',
\newblock in {\em Proc. 48th Annual Allerton}, Monticello, IL, 2010.


\bibitem{DRWS10}
A.~G. Dimakis, K.~Ramchandran, Y.~Wu, and C.~Suh,
\newblock ``A survey on network codes for distributed storage'',
\newblock {\em Proc. of the IEEE}, vol. 99, no. 3, March 2011.








\bibitem{me}
T.~Ernvall, S.~El Royhayeb, C.~Hollanti, and V.~Poor,
\newblock ``Heterogeneous distributed storage systems: capacity and security
  results'',
\newblock {\em J. on Selected Areas in Communications}, Dec. 2013.


\bibitem{rashmietal}
K.V. Rashmi, N.B. Shah, and P.V. Kumar,
\newblock ``{Optimal Exact-Regenerating Codes for Distributed Storage at the
  MSR and MBR Points via a Product-Matrix Construction}'',
\newblock {\em IEEE Trans. Inf. Theory}, vol. 57, no. 8, August
  2011.



\bibitem{Ning}
N.~Wang and J.~Lin,
\newblock ``{Joint Channel-Network Coding (JCNC) for Distributed Storage in
  Wireless Network}'',
\newblock {\em Lecture Notes of the Institute for Computer Sciences, Social
  Informatics and Telecommunications Engineering}, vol. 4, 2009.

\bibitem{gong}
C.~Gong,
\newblock ``{On Partial Downloading for Wireless Distributed Storage
  Networks}'',
\newblock {\em IEEE Trans. on Signal Processing}, vol. 60,
June 2012.


\bibitem{rashmi_erasure} K. V. Rashmi, N. B. Shah, K. Ramchandran, P. V. Kumar, ``Regenerating codes for errors and erasures in distributed storage'', \emph{IEEE ISIT 2012}, July 2012.









\bibitem{oggierviterbo2}
F.~Oggier, E.~Viterbo, and J.-C. Belfiore,
\newblock ``{Cyclic Division Algebras: A Tool for space--time Coding}'',
\newblock {\em Foundations and Trends in Communications and Information
  Theory}, vol. 4, no. 1, 
  2007.












\bibitem{ZT-MAC} D. N. C. Tse, P. Viswanath, and L. Zheng, ``Diversity-multiplexing
tradeoff in multiple-access channels'', \emph{IEEE Trans. Inf. Theory}, 50(9), 
Sep. 2004.


\bibitem{kuser}	 H.-F. Lu, C. Hollanti, R. Vehkalahti, and J. Lahtonen, ``DMT optimal code constructions 	for multiuser MIMO channel'', \emph{IEEE Trans. Inf. Theory},  57(6),  June 2011. 












 








\bibitem{Lusina} 
P. Lusina, E. Gabidulin, and M. Bossert, ``Maximum rank distance codes as space--time codes'', {\em IEEE Trans. Inf. Theory}, 49(10), 2003.

\bibitem{Lu-Kumar}
H.-F. Lu and P.V. Kumar, ``Unified construction of space--time codes with optimal rate-diversity tradeoff'', {\em IEEE Trans. Inf. Theory}, 51(5), 2005. 

\bibitem{BB} M. Badr and J.-C. Belfiore,  ``Distributed space--time block codes for the non cooperative multiple access channel,'' \emph{2008 IEEE Int. Zurich Seminar on Comm.}, pp.132--135, March 2008.




\bibitem{remarks}	H.-F. Lu, J. Lahtonen, R. Vehkalahti, and C. Hollanti, ``Remarks on the criteria of constructing MIMO-MAC DMT optimal codes'', \emph{2010 IEEE Inf. Theory Workshop (ITW)}, Cairo, 2010. 




\bibitem{CNVD} C. Hollanti, H.-F. Lu, and R. Vehkalahti, ``An algebraic tool for obtaining
conditional non-vanishing determinants'', \emph{2009 IEEE Int. Symp. on Inf. Theory},  Seoul, 2009.


\end{thebibliography}
\end{document}